\documentstyle{aipproc}
\newcommand{\bq}{\begin{equation}}
\newcommand{\eq}{\end{equation}}

\newcommand\ka{\kappa_1}
\newcommand\kb{\kappa_2}
\newcommand\kap{\kappa_1'}
\newcommand\kbp{\kappa_2'}
\newcommand\xx{\tilde{x}}
\newcommand\AAA{\alpha_1}
\newcommand\AB{\alpha_2}
\newcommand\Kvec{\mbox{\boldmath $K$}}

\newcommand\FFA{\mbox{$\widetilde{F}^s$}}

\begin{document}

\title{
Evolution Kernels for Light-Ray\\ Operators~: \\
Twist~2 and Twist~3 Contributions}

\author{Johannes Bl\"umlein$^*$, Bodo Geyer$^{\dagger}$
 and Dieter Robaschik$^*$}
\address{$^*$DESY -- Zeuthen, Platanenallee 6, D -- 15735 Zeuthen,
 Germany \\
$^{\dagger}$
Naturwissenschaftlich--Theoretisches Zentrum der
Universit\"at Leipzig,\\
Augustusplatz 10, D--04109 Leipzig, Germany}

\maketitle

\begin{abstract}
The general evolution kernels of the twist~2 light--ray operators for
unpolarized and polarized deep inelastic scattering are calculated in
${\cal O}(\alpha_s)$. From these evolution kernels a series of special
evolution equations can be derived, among them the Altarelli-Parisi
equations and the evolution equation for the meson wave function.
In the case of twist 3 the results of Balitzki and Braun are confirmed.
\end{abstract}

\section*{Introduction}
%
%
%
%

\noindent
The study of the Compton amplitude for scattering of a virtual photon
off a hadron is one of the basic tools in QCD to understand the
short--distance behavior of the theory.
The Compton amplitude for the general
case of  non--forward scattering is given by
\begin{equation}
\label{E:JBL:COMP}
T_{\mu\nu}(p_+,p_-,Q) = i \int d^4x e^{iqx}
\langle p_2|T (J_{\mu}(x/2) J_{\nu}(-x/2))|p_1\rangle,
\end{equation}
where $p_+ = p_2 + p_1,~p_- = p_2 - p_1 = q_1 - q_2$ and
$q = (q_1 + q_2)/2$.
The time--ordered product  in eq.~(\ref{E:JBL:COMP}) can be
represented in terms of the operator product expansion. Here we use
the
representation derived in ref.~\cite{JBL:AS,JBL:LEIP}
\begin{eqnarray}
\label{tpro}
T (J_{\mu}(x/2) J_{\nu}(-x/2)) &\approx&
\int_{-\infty}^{+\infty}  d \kappa_-
\int_{-\infty}^{+\infty}  d \kappa_+
 [ C_a(x^2, \kappa_i,  \mu^2) {S_{\mu\nu}}^{\rho\sigma}
\xx_{\rho} O_{\sigma}^a (\kappa_i\xx, \mu^2) \nonumber
\\
& &
+ C_{a,5}(x^2, \kappa_i, \mu^2)
{\varepsilon_{\mu\nu}}^{\rho\sigma}
\xx_{\rho} O_{5, \sigma}^a (\kappa_i\xx, \mu^2) ],
\end{eqnarray}
with $S_{\mu\nu\rho\sigma} = g_{\mu\nu} g_{\rho\sigma}
- g_{\mu\rho} g_{\nu\sigma} - g_{\mu\sigma} g_{\nu\rho}$ and
$\varepsilon_{\mu\nu\rho\sigma}$ denoting the Levi--Civita symbol.
The light--like vector
$\xx = x + r (x.r/r.r) \left [ \sqrt{1 - x.x r.r / (x.r)^2} - 1 \right ]
$ is related to $x$ and a subsidiary four--vector
$r$, $C_a$ and  $C_{a5}$ denote the respective coefficient functions.
Different kinematic situations in the Bjorken region are described
by different matrix--elements of the involved light--ray operators
$ O_{\sigma}^a (\kappa_+ \xx, \kappa_- \xx, \mu^2)$ and
$ O_{5, \sigma}^a (\kappa_+ \xx, \kappa_- \xx, \mu^2)$.
Let us illustrate this with the help of the non--singlet
quark operator,  given in
the axial gauge $\tilde x_{\mu} A^{\mu}(x) =0$ by
\begin{eqnarray}
O^{\rm NS}(\ka,\kb) =
i
\left [
\overline{\psi_a}(\ka\xx)
\lambda_f
\gamma_{\mu} \xx^{\mu} \psi_a(\kb \xx) \right ].
\end{eqnarray}
For deep inelastic forward scattering, $p_1=p_2=p$, the parton
distribution reads\footnote{For brevity we put $\kappa_1 = -\kappa_-,
\kappa_2 = \kappa_- \equiv \kappa$ in eqs.~(\ref{E:JBL:4}--\ref{E:JBL:7}).
}
\begin{eqnarray}
\label{E:JBL:4}
q(z,\mu^2) = \int_{-\infty}^{+\infty}
 \frac{d(\tilde x p\kappa)}{2\pi \tilde x p}
             e^{2i\kappa \tilde x p z }
    \langle p|\overline{\psi}_a(-\kappa \tilde x)\lambda_f
\gamma \tilde x
              \psi_a(+\kappa\tilde x)|p \rangle
\end{eqnarray}
and obeys the Altarelli--Parisi equation.
The meson wave function, on the other hand,
 has the representation
\begin{eqnarray}
\Phi(t,\mu^2) = \int_{-\infty}^{+\infty}
\frac{d(\tilde x p\kappa)}{2\pi \tilde x p}
             e^{i\kappa \tilde x p t }
    \langle 0|\overline{\psi}_a(-\kappa \tilde x)\lambda_f
\gamma \tilde x
              \psi_a(+\kappa\tilde x)|p \rangle
\end{eqnarray}
and satisfies the Brodsky--Lepage equation.
For general non--forward processes the distribution function is
defined \cite{JBL:LEIP} by
\begin{eqnarray}
F(z_+,z_-,\mu^2)\! &=& \!\!
{\int \!\!\!\!
\int}_{-\infty}^{+\infty}
\frac{d(\kappa\tilde x p_+)}
{2\pi \tilde x p_+} \frac{d(\kappa\tilde x p_-)}
{2\pi }
             e^{i\kappa \tilde x p_+ z_+ +i\kappa\tilde x p_-z_-}
\\ & &~~~~~~~~~~~~~~~~~~~~~~~~~\times
  \langle p_2|\overline{\psi}_a(-\kappa \tilde x)\lambda_f
\gamma \tilde x
              \psi_a(+\kappa\tilde x)|p_1 \rangle.
\nonumber
\end{eqnarray}
The corresponding evolution equations are given by
 eq.~(\ref{JBL:evo3}), see also \cite{JBL:RAD}.
If we restrict the general non--forward process by the condition
$\tau =
\tilde x p_- /\tilde x p_+ $, $|\tau| \leq 1 $, then a
one-variable distribution function \cite{JBL:LEIP}
\begin{eqnarray}
\label{E:JBL:7}
q(t, \tau, \mu^2) &=& \int_{-\infty}^{+\infty}
 \frac{d(\tilde x p_+\kappa)}{2\pi \tilde x p_+}
             e^{i\kappa \tilde x p_+ t }
             \langle p_2|\overline{\psi}_a(-\kappa 
\tilde x)\lambda_f
\gamma \tilde x
        \psi_a(+\kappa\tilde x)|p_1\rangle|_{\tilde x p_-
= \tau \tilde x p_+}\nonumber\\
          &= & \int_{-\infty}^{+\infty} dz_- F(t-\tau z_-,z_-).
\end{eqnarray}
is obtained.
Note that the partition functions $q(z,\mu^2)$ and $\Phi(t,\mu^2)$
 can be obtained as limits
$\tau \rightarrow 0$,
$\tau = \pm 1$, respectively, see ref.~\cite{JBL:BGR1}.

\section*{The Evolution Kernels}
\noindent
A consequence of the relation of different distribution functions to
a single
operator is, that the evolution kernel, i.e. the anomalous dimension,
 of this operator allows
the derivation of the corresponding kernels which emerge in a variety
of different processes. On the other hand,
it underlines the importance to determine the general
kernels for all relevant operators themselves. In the following
we consider the flavor singlet operators only\footnote{The
non--singlet operators are discussed in \cite{JBL:BGR1}.}.
\begin{eqnarray}
O^q_{\rm     }(\ka,\kb) &=&
\frac{i}{2} \left [
\overline{\psi_a}(\ka\xx)
\gamma_{\mu} \xx^{\mu} \psi_a(\kb \xx) -
\overline{\psi_a}(\kb\xx)
\gamma_{\mu} \xx^{\mu} \psi_a(\ka \xx) \right ]\\
O^q_{\rm 5   }(\ka,\kb) &=&
\frac{i}{2}
\left [
\overline{\psi_a}(\ka\xx) \gamma_5
\gamma_{\mu} \xx^{\mu} \psi_a(\kb \xx) +
\overline{\psi_a}(\kb\xx) \gamma_5
\gamma_{\mu} \xx^{\mu} \psi_a(\ka \xx) \right ]\\
O^G_{\rm     }(\ka,\kb) &=&
\xx^{\mu} {F_{s\mu}}^{\nu}(\ka\xx) \xx^{\mu'} {F^s}_{\mu'\nu}(\kb\xx)\\
O^G_{\rm 5   }(\ka,\kb) &=& \frac{1}{2} \left [
\xx^{\mu} {F_{s \mu}}^{\nu}(\ka\xx) \xx^{\mu'}
\FFA_{\mu'\nu}(\kb\xx)
- \xx^{\mu} {F^s}_{\mu\nu}(\kb\xx) \xx^{\mu'}
{\FFA_{\mu'}}{\vspace*{-1.2mm}^{\nu}}(\ka\xx)
\right ], \nonumber\\
\end{eqnarray}
where $\psi_a$ denotes the quark
and $F_{\mu\nu}^s$ the gluon field strength operators, respectively.
The operator dual
to $F_{\mu\nu}^s$ is
$\widetilde{F}_{\mu\nu}^s = \frac{1}{2}
\varepsilon_{\mu\nu\rho\sigma} F^{s \rho\sigma}$,
and $\kappa_1 = \kappa_+ - \kappa_-,~\kappa_2
= \kappa_+ + \kappa_-$.

The renormalization group equation implies
the following evolution equations for these operators~:
\begin{eqnarray}
\label{evo1}
\mu^2 \frac{d}{d \mu^2} \left (\begin{array}{c}
O^q(\kappa_i) \\
\label{evo2}
O^G(\kappa_i) \end{array} \right ) &=&
\frac{\alpha_s(\mu^2)}{2 \pi}
\int_0^1 d \AAA \int_0^{1-\AAA} d \AB
\Kvec(\AAA,\AB)
\left (\begin{array}{c}
O^q(\kappa'_{i}) \\
O^G(\kappa'_{i}) \end{array} \right ),
\nonumber \\
\end{eqnarray}
with $\alpha_s = g_s^2/(4\pi)$ the strong coupling constant, $\mu$
the renormalization scale, and  $\kap = \ka(1-\AAA)+\kb\AAA,
\kbp = \kb(1-\AB)+\ka\AB$.  $\Kvec$
denotes the
matrix of the singlet
evolution kernels in  the unpolarized case. The singlet evolution
equations for the polarized case are obtained replacing $O^{q,G}$ by
$O^{q,G}_{5}$ and $\Kvec$ by $\Delta \Kvec$. As far as relations are
concerned which are valid both for the unpolarized and polarized case
under this replacement, we will only give that for the
unpolarized case  in the following. The matrices of the
singlet kernels are
\begin{equation}
\Kvec = \left ( \begin{array}{ll} K^{qq} & K^{qG} \\
                                  K^{Gq} & K^{GG} \end{array} \right )
~~~~{\rm and}~~~~\Delta \Kvec = \left ( \begin{array}{ll} \Delta K^{qq} &
\Delta K^{qG} \\ \Delta K^{Gq} & \Delta K^{GG} \end{array} \right )~,
\end{equation}
respectively.
In the unpolarized case the kernels are given by
\begin{eqnarray}
\label{E:JBL:eqK1}
K^{qq}(\AAA,\AB) &=&  C_F  \Biggl\{ 1 - \delta(\AAA) - \delta(\AB)
+ \delta(\AAA) \left [ \frac{1}{\AB}\right]_+
+ \delta(\AB) \left [ \frac{1}{\AAA}\right]_+ \\\nonumber
& &+ \frac{3}{2} \delta(\AAA)\delta(\AB)  \Biggr
\}\\
K^{qG}(\AAA,\AB) &=&  - 2 N_f T_R \kappa_-
\left \{ 1 - \AAA - \AB + 4 \AAA\AB
\right
\} \\
K^{Gq}(\AAA,\AB) &=&  - C_F \frac{1}{\kappa_- - i \varepsilon}
\left \{  \delta(\AAA) \delta(\AB) + 2 \right \} \\
\label{E:JBL:eqK4}
K^{GG}(\AAA,\AB) &=&  C_A  \Biggl
\{ 4 ( 1 - \AAA -\AB) + 12 \AAA \AB
+\delta(\AAA) \left (
 \left [\frac{1}{\AB} \right ]_+ - 2
+ \AB \right )\\
& & + \delta(\AB) \left ( \left [\frac{1}{\AAA} \right ]_ +        - 2
+ \AAA \right )   \Biggr
\}
+ \frac{\beta_0}{2} \delta(\AAA)\delta(\AB),   \nonumber
\end{eqnarray}
in ${\cal O}(\alpha_s)$,
where $C_F = (N_c^2-1)/2N_c \equiv 4/3, T_R = 1/2, C_A = N_c \equiv 3$,
$\beta_0 = (11 C_A - 4 T_R N_f)/3$,
$N_f$  being
the number of active quark flavors.
The $+$-prescription is defined as
\begin{equation}
\int_0^1 dx
 \left [f(x,y)\right]_+ \varphi(x) = \int_0^1 dx
f(x,y)
\left [\varphi(x) - \varphi(y) \right]~,
\end{equation}
where the singularity of $f$ is of the type $\sim 1/(x - y)$.

The  kernels for the polarized case are
\begin{eqnarray}
\label{E:JBL:eqDK1}
\Delta K^{qq}(\AAA,\AB) &=& K^{qq}(\AAA,\AB) \\
\Delta K^{qG}(\AAA,\AB) &=&  - 2 N_f T_R \kappa_-
\left \{ 1 - \AAA - \AB \right \} \\
\Delta K^{Gq}(\AAA,\AB) &=&  -  C_F \frac{1}{\kappa_- - i \varepsilon}
\left \{\delta(\AAA) \delta(\AB) - 2 \right \} \\
\label{E:JBL:eqDK4}
\Delta K^{GG}(\AAA,\AB) &=&  K^{GG}(\AAA,\AB) - 12 C_A   \AAA \AB~.
\end{eqnarray}
Whereas the kernels for the polarized case have been
derived for the first
time in ref.~\cite{JBL:BGR1} recently,
those for the unpolarized case were found several years ago
already in refs.~\cite{JBL:BGR,JBL:BB}.
The kernels $K^{ij}$ and $\Delta K^{ij}$ determine the respective
evolutions of the operators
$O^{q}_{(5)}$, and $O^{G}_{(5)}$ in ${\cal O}(\alpha_s)$.

The evolution kernels,
 eq.~(\ref{E:JBL:eqK1}--\ref{E:JBL:eqK4}) and
 eq.~(\ref{E:JBL:eqDK1}--\ref{E:JBL:eqDK4}),
determine the evolution of all
physically
interesting partition functions. As an example we
consider the general
two--variable partition functions, introduced in the introduction,
which are
defined by
\begin{eqnarray}
\frac{\langle p_1|O^{q}|p_2\rangle}{(i\xx p_+)} &=&
e^{-i\kappa_+ \xx p_-}
\int_{-\infty}^{+\infty} d z_+
\int_{-\infty}^{+\infty} d z_-
e^{-i\kappa_-( \xx p_+ z_+ + \xx p_- z_-)} F_{q}(z_-,z_+),\\
\frac{\langle p_1|O^{G}|p_2\rangle}{(i\xx p_+)^2} &=&
e^{-i\kappa_+ \xx p_-}
\int_{-\infty}^{+\infty} d z_+
\int_{-\infty}^{+\infty} d z_-
e^{-i\kappa_-( \xx p_+ z_+ + \xx p_- z_-)} F_{G}(z_-,z_+)~.
\end{eqnarray}
The singlet
 evolution equations for these functions read
\begin{eqnarray}
\label{JBL:evo3}
\mu^2 \frac{d}{d \mu^2}
\left (\begin{array}{c}
F^{q}(z_+,z_-) \\
F^{G}(z_+,z_-) \end{array}\right )
&=&
\frac{\alpha_s(\mu^2)}{2 \pi}
\int_{-\infty}^{+\infty} \frac{d z'_+}{|z'_+|}
\int_{-\infty}^{+\infty} d z'_- \\
& &~~~~~~~~~~~~~\widetilde{\Kvec}(z_+,z_-;z'_+,z'_-)
\left (\begin{array}{c}
F^{q}(z'_+,z'_-) \\
F^{G}(z'_+,z'_-) \end{array}\right ), \nonumber
\end{eqnarray}
where
\begin{eqnarray}
\lefteqn{\tilde{K}^{ij}(\alpha_1,  \alpha_2) = } \\
& &~~~~~~~\frac{1}{2} \int_0^1 dz''_+ \widetilde{O}^{ij}(z_+,z''_+)
 \theta(1 - \alpha'_+) \theta(\alpha'_+)
 \theta(\alpha'_+  +  \alpha'_-)
 \theta(\alpha'_+  -  \alpha'_-)
K^{ij}(\alpha'_1,  \alpha'_2)~\nonumber,
\end{eqnarray}
with $\alpha'_{\rho} =  \alpha_{\rho}(z_+ \rightarrow z''_+)$,
and
$\alpha_1 = (\alpha_+ + \alpha_-)/2,~~~~
\alpha_2 = (\alpha_+ - \alpha_-)/2,$
$\alpha_+ = 1 - z_+/z'_+,~~~~\alpha_-
= (z_+ z'_- - z_- z'_+)/z'_+ $,
and
\begin{equation}
\widetilde{O}^{ij}(z_+,z''_+) =
\left( \begin{array}{rr} \delta(z_+ - z''_+)&
  \partial_{z_+}  \delta(z_+ - z''_+)\\
- \theta(z_+ - z''_+)  &\delta(z_+ - z''_+) \end{array}
\right )~.
\end{equation}
The determination of the corresponding Brodsky-Lepage and
Efremov-Radyushkin kernels, the extended kernels, as well as the
Altarelli-Parisi limiting case and the relation to other
investigations \cite{JBL:RAD,JBL:XJ,JBL:PEN} are discussed in
ref.~\cite{JBL:BGR1}.

\section*{Twist 3}
\noindent
The consideration of the renormalization properties of the twist 3
operators is much more complicated. The reason is that several operators
exist which are related by each other due to
the equations of motion. If one
introduces the Shuryak--Vainshtein operators and applies
the equations of motion consequently, then
results by Balitzki
and Braun, as well as by Bukhvosov et al.
and others are reproduced. These results were already published in 
ref.~\cite{JBL:GMR}.
However, this  may yet  not be the complete solution of the
problem and further investigations are needed.

\vspace{2mm}
\noindent
{\bf Acknowledgement.}
We would like to thank Paul S\"oding for his
constant support of the  project and D. M\"uller for discussions on
the present topic.

\end{document}